\documentclass[aip,cha,reprint,amsmath,amssymb,raggedbottom,nobalancelastpage,sort]{revtex4-2}

\usepackage{graphicx}
\usepackage{bm}
\usepackage{xurl}
\usepackage[colorlinks=true,linkcolor=blue,citecolor=blue,urlcolor=blue]{hyperref}

\DeclareMathOperator{\erfc}{erfc}

\begin{document}

\title{Comment on ``Note on the start-up of Couette flow for viscoelastic fluids'' [Phys.\ Fluids \textbf{35}, 113108 (2023)]}

\author{Ivan C. Christov}
\email{christov@purdue.edu}
\homepage{https://christov.tmnt-lab.org}
\affiliation{School of Mechanical Engineering, Purdue University, West Lafayette, Indiana 47907, USA}

\date{\today}

\begin{abstract}
I show that the initial conditions imposed by Balan [Phys.\ Fluids \textbf{35}, 113108 (2023)] are incompatible with a nonzero retardation time, so that the problem solved is not impulsive start-up, but, as I prove exactly, the start-up of a plate ramped as $1 - \mathrm{e}^{-t/\lambda_2}$ on the retardation timescale $\lambda_2$ (a material parameter, not a setting in a rheometer). Values digitized from that paper's viscoelastic figures fall on the ramped-plate solution, not on Tanner's start-up solution. The large negative wall normal stress reported for the corotational model is a grid-dependent artifact due to the same error in the initial data. An open-source code repository provides annotated notebooks reproducing every figure and number in this Comment.
\end{abstract}

\maketitle

In a recent paper, Balan~\cite{balan2023} solved the unidirectional-flow momentum equation numerically under a three-constant Gordon--Schowalter~\cite{gordon_schowalter1972,johnson_segalman1977} (Jeffreys-type) constitutive relation for a viscoelastic fluid [Eqs.~(5)--(7) of Ref.~\onlinecite{balan2023}], for the start-up of plane Couette flow. The stated aim was to extend Tanner's classical solution~\cite{tanner1962} of Stokes' first problem for an Oldroyd-B fluid to the corotational (Jaumann) model, whose steady flow curve is non-monotonic. In this Comment, I show that the initial-boundary-value problem (IBVP) posed in Eqs.~(9)--(10) of Ref.~\onlinecite{balan2023} is \emph{not} the impulsive start-up problem. I identify the problem actually solved and quantify its consequences for the results of Ref.~\onlinecite{balan2023}. I also show that the steady states reported there are not converged in space, and correct several statements made therein about the earlier literature.

In the notation of Ref.~\onlinecite{balan2023}, with $\kappa = \lambda_2/\lambda_1$ the ratio of retardation to relaxation time, the dimensionless governing equations are
\begin{subequations}\label{eq:1}
\begin{align}
    \mathrm{Re}\,\partial_t v &= \partial_x \sigma, \label{eq:1a}\\
    \partial_t N + N &= 2(1-a^2)\big[\sigma\partial_x v - \kappa(\partial_x v)^2\big], \label{eq:1b}\\
    \partial_t\sigma + \sigma &= \kappa\partial_x\partial_t v - (N/2 - 1)\partial_x v, \label{eq:1c}
\end{align}
\end{subequations}
which I verified symbolically. Velocities are scaled by the plate speed $v_0$, so $v_0=1$ in these units. The symbol is retained below, as in Ref.~\onlinecite{balan2023}, only to mark the amplitude. The initial conditions (ICs) imposed in Ref.~\onlinecite{balan2023} are $v(x,0) = v_0 H(x - x_0)$ [Eq.~(9)(iii)] and $\sigma(x,0) = 0$ [Eq.~(10)(iv)], where $x_0$ is the position of the moving plate, written $\bar x$ in Ref.~\onlinecite{balan2023}, since an overbar here denotes the Laplace transform in $t$ throughout. The \texttt{NDSolve} calls reproduced in Annex~B of the supplementary material of Ref.~\onlinecite{balan2023} confirm that these are the conditions passed to the solver (``\texttt{tau[x, 0] == 0}'' together with ``\texttt{v[x, 0] == UnitStep[x - 100]}''). 

For $\kappa>0$, these conditions \emph{cannot} both hold at the same instant. The IC $v(x,0)=v_0H(x-x_0)$ is the state at $t=0^+$, since at $t=0^-$ the fluid is at rest. Meanwhile, the IC $\sigma(x,0)=0$ is the state at $t=0^-$. Write $\sigma=S+\kappa\partial_xv$, where $S$ is the polymer contribution and $\kappa\partial_xv$ is the instantaneous (solvent, or retardation) one. In these variables no $\partial_tv$ remains in Eq.~\eqref{eq:1c}, $S$ cannot jump at $t=0$, and $S(x,0^-)=0$ (an unstressed fluid) gives $S(x,0^+)=0$. Since $\partial_xv(x,0^+)=v_0\delta(x-x_0)$, it follows that $\sigma(x,0^+)=\kappa v_0\delta(x-x_0)\neq0$. Imposing $\sigma(x,0)=0$ at $t=0^+$ therefore prescribes a fictitious polymer stress $S(x,0^+)=-\kappa v_0\delta(x-x_0)$, which subsequently relaxes as $\mathrm{e}^{-t}$. The consistently posed problem instead prescribes the state before start-up, i.e., $v(x,0^-)=S(x,0^-)=0$ and $N(x,0^-)=N_0$, and leaves the jump to the boundary conditions, $v(x_0,t)=v_0H(t)$ and $v(0,t)=0$. Clearly, $\sigma$ needs no IC at all, as it follows from $S$ and $\partial_xv$ at any $t$.

As shown in this Comment's Appendix, for $a = \pm 1$, IBVP (9)--(10) is exactly the consistently posed start-up problem for a plate whose velocity is $v_0(1 - \mathrm{e}^{-t/\kappa})$ in the units of Ref.~\onlinecite{balan2023}, i.e., $v_0(1 - \mathrm{e}^{-t/\lambda_2})$ in dimensional form. The initial data in Ref.~\onlinecite{balan2023} replace the step with an exponential ramp on the retardation timescale. This is the rate-type counterpart of an existing error for a second-grade fluid~\cite{christov_christov2010}: there, the claimed start-up solution is actually a known ramp-up solution~\cite{barenblatt1960}. In both cases, the problem actually solved is well posed, but it is \emph{not} the start-up problem, which is what makes the error easy to miss and, once suspected, easy to diagnose. 

Importantly, this point is \emph{rheological} rather than numerical: a nonzero retardation time is the statement that the fluid has an instantaneous, solvent-like response~\cite{oldroyd1950}, so a step in the wall velocity must be accompanied by a singular stress at $t = 0^+$. Imposing zero initial stress denies the fluid that response.

\begin{figure*}
\includegraphics[width=\textwidth]{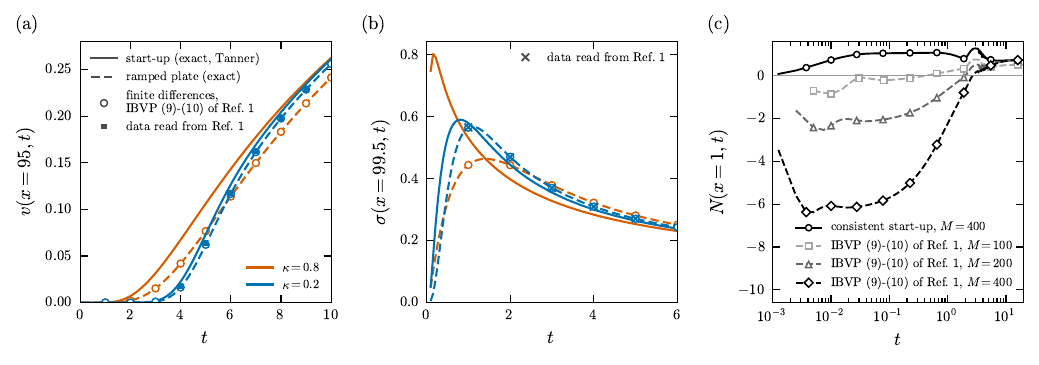}
\caption{Wide gap ($x_0=100$ in units of the elastic length $\ell=\sqrt{\eta_0\lambda_1/\rho}$, effectively a half-space): (a) $v(x = 95, t)$ and (b) $\sigma(x = 99.5, t)$, both for $a = 1$, $\mathrm{Re} = 1$ and $\kappa = 0.8$ and $0.2$. Curves are exact solutions: Tanner's start-up problem [Ref.~\onlinecite{christov_jordan2009}, Eq.~(7)] (solid) and the ramped-plate problem from the Appendix (dashed). Symbols are numerical: finite-difference solution of IBVP (9)--(10) as posed in Ref.~\onlinecite{balan2023} (open) and values read from Fig.~3(a) of Ref.~\onlinecite{balan2023} (filled squares in (a), crosses in (b)), which exist for $\kappa = 0.2$ only [see remark~(iv) below]. (c) Unit gap, Couette ($x_0=1$ in units of $h$, so $h=\ell$ at $\mathrm{Re}=1$): normal stress at the wall, $N(1,t)$, for $a = 0$, $\kappa = 0.01$, $\mathrm{Re} = 1$, $N_0 = 0$, finite differences: consistently posed start-up (solid) and IBVP (9)--(10) of Ref.~\onlinecite{balan2023} on different grids (dashed).}
\label{fig:1}
\end{figure*}

Panels (a) and (b) of Fig.~\ref{fig:1} are computed at the retardation time of Fig.~3(a) of Ref.~\onlinecite{balan2023} and at a larger one, which Ref.~\onlinecite{balan2023} does not compute but where the discrepancy is largest. The exact curves are evaluated using Eq.~(7) of Ref.~\onlinecite{christov_jordan2009}, with the breakpoints of its Eq.~(8) and $\beta = 0$; Eq.~(7) is the Laplace inverse~\cite{tanner1962} of $\bar v(x,p) = p^{-1}\exp[-(x_0-x)\sqrt{p(1+p)/(1+\kappa p)}]$ at $\mathrm{Re}=1$, and the ramped-plate solution follows from Duhamel's principle~\cite{christov2013}. The finite-difference solutions are second order, with the trapezoidal rule (Crank--Nicolson) in time. With consistent initial data, they reproduce the exact step solution on a grid with 5000 cells to better than $5\times10^{-7}$ in $v$ for (a) and $4\times10^{-6}$ in $\sigma$ for (b)~\cite{christov_code2026}.

The values read from Figs.~3(a) and 4(a) of Ref.~\onlinecite{balan2023} do not agree with Tanner's solution. During the transient, they fall short by roughly a fifth for $\kappa = 0.2$ (the case of Fig.~\ref{fig:1}) and by up to a third for $\kappa = 0.4$. They agree instead with the ramped-plate curve and with the finite-difference solution of IBVP (9)--(10) as posed. The same holds for the near-wall stress peak of Fig.~3(a) of Ref.~\onlinecite{balan2023}: the ramped-plate problem's peak is both lower and later than Tanner's, and the values read from Ref.~\onlinecite{balan2023} follow it [Fig.~\ref{fig:1}(b)]. The Newtonian curves of Figs.~2 and 4(a) of Ref.~\onlinecite{balan2023}, by contrast, agree with $\erfc[(x_0-x)/(2\sqrt t)]$, Stokes' solution at $\mathrm{Re}=1$, which suggests that they were obtained differently. The comparison with Tanner~\cite{tanner1962} in Fig.~B.1 of the supplementary material of Ref.~\onlinecite{balan2023} uses ``some values extracted for the Tanner's graphs'' [sic] and reports ``a fair correspondence''. It is not fine enough to detect the discrepancy, so the statement in Ref.~\onlinecite{balan2023} that ``Solutions at $\kappa = 0.2$ and $\kappa = 0.4$ are consistent with the results from Tanner's paper'' does not hold at the level of the curves themselves.

For $\kappa = 0.01$, the value used for the corotational model in Ref.~\onlinecite{balan2023}, the direct effect is small for $a = \pm 1$. For $t \ge 0.5$, the wall shear stress changes by about 2\%. For $a = 0$, however, IBVP (9)--(10) changes the transient by an amount that depends on the grid spacing: up to 0.18, 0.06, and 0.03 in $v$ using 400, 1000 (the grid of Ref.~\onlinecite{balan2023}), and 1600 cells, and 0.05 (11\%) in the wall shear stress near $t = 3.5$--$4$ on the 1000-cell grid. The steady state is unaffected. The mechanism is that, for $a = 0$, the fictitious stress $S(x,0)$ multiplies the initial shear rate $v_0 \delta(x - x_0)$ in Eq.~\eqref{eq:1b}, which in terms of $S$ reads $\partial_t N + N = 2(1-a^2)S\,\partial_x v$ (the solvent part of $\sigma$ cancels), producing a negative source in the normal stress of magnitude $\sim 2\kappa v_0^2/\Delta x^2$ that acts over the transit time of the elastic wave across one cell, $\sim\Delta x/c=\Delta x$ at $\mathrm{Re}=1$, where $\Delta x$ is the grid spacing and $c$ is the wave speed.

Figure~\ref{fig:1}(c) shows that a large negative wall normal stress, such as the $N(1,t) \simeq -4$ of Fig.~8 of Ref.~\onlinecite{balan2023}, is a property of the discretization and not of the model. For IBVP (9)--(10), the minimum of $N$ at the wall is $-0.85$, $-2.5$, and $-6.4$ on grids of 100, 200, and 400 cells, and grows without bound as the grid is refined, whereas for the consistently posed start-up $N$ never goes negative, its minimum on the 400-cell grid being $0.09$. Since the dependence of the steady state on the initial normal stress is the principal conclusion of Ref.~\onlinecite{balan2023}, this transient is not a side issue. The stress-controlled simulations of Ref.~\onlinecite{balan2023} (Figs.~13 and 14), which start from $v = \sigma = 0$, have consistent initial data and are not affected. More generally, start-up transients are where rate-type models differ most~\cite{divoux2016,stephanou2023} (stress overshoot, approach to a non-monotonic flow curve, shear banding), so initial data that are inconsistent with a model's instantaneous response perturb precisely the quantity under study.

The steady states of Ref.~\onlinecite{balan2023} are also \emph{not} converged in space. Annex~B of the supplementary material of Ref.~\onlinecite{balan2023} reports that, for the corotational simulations, \texttt{NDSolve} returned a ``scaled local spatial error estimate'' of 199 on the 1001-point grid used for the paper and of 89 on a 5001-point grid, ``much greater than the prescribed error tolerance,'' adding that ``a singularity may have formed.'' It also reports that the steady wall shear stress differs by 0.8\% from its average across the gap, and that on finer grids the kink moves toward the plate and $\sigma_0$ increases. Thus, by the solver's own estimate, the steady states in Figs.~7--9 of Ref.~\onlinecite{balan2023} are not converged in space, and the oscillations in those figures and the hatched ``domains of possible variations'' in Fig.~7 are discretization effects, as is the $\kappa = 0.001$ stress curve of Fig.~9, which is not uniform across the gap as Eq.~\eqref{eq:1a} requires of any steady state. In the present simulations, the wall-stress oscillations decrease monotonically as the grid is refined from 100 to 3200 cells, with $\sigma_0$ settling at 0.482. This is in line with the unsteady finite-element solutions of the same problem (Johnson--Segalman model, Newtonian solvent, and inertia) by Georgiou and Vlassopoulos~\cite{georgiou1998}, who found that the final shear stress and the number and location of the kinks are set by the initial perturbation. 

Four final remarks are in order. (i)~The choice $\mathrm{De}=1$ in Ref.~\onlinecite{balan2023} is not ``without loss of generality.'' The Deborah number, $\mathrm{De}$, appears explicitly in the dimensionless constitutive relation [Eq.~(a.6) of Annex~A of the supplementary material] and multiplies $\partial_x v$ in Eqs.~\eqref{eq:1b} and \eqref{eq:1c}. The underlying issue is the nondimensionalization. For unidirectional flow, the convective acceleration vanishes identically, so no Reynolds number arises. Scaling velocity by $v_0$, length by $h$, and time by $\lambda_1$ \emph{manufactures} two groups, $\mathrm{Re}$ and $\mathrm{De}$, and once $\mathrm{De}$ is fixed the ``Reynolds number'' of Ref.~\onlinecite{balan2023} is $\rho h^2/(\eta_0\lambda_1) = (h/\ell)^2$, with $\ell = \sqrt{\eta_0\lambda_1/\rho}$ the elastic length that Tanner~\cite{tanner1962} used as the length scale (hence ``$\mathrm{Re}=1$ in (8)'' in Ref.~\onlinecite{balan2023}). 

With $\ell$, $\lambda_1$, the elastic wave speed~\cite{joseph1986} $c=\ell/\lambda_1$, and $\eta_0/\lambda_1$ as the scales, Eq.~\eqref{eq:1a} becomes $\partial_t v = \partial_x\sigma$, Eqs.~\eqref{eq:1b} and \eqref{eq:1c} are unchanged, the gap is $0<x<\Lambda=h/\ell$, and the plate moves at $v_0/c$, $v_0$ now being dimensional. The linear problem ($a=\pm1$) then depends on $\Lambda$ and $\kappa$ only, with $v_0$ scaling out. The corotational problem adds $v_0/c$, i.e., the imposed shear rate $\lambda_1 v_0/h$, which is the only place $\mathrm{De}$ enters. At $\mathrm{De}=1$ and $\kappa=0.01$, that shear rate sits at the maximum of the first stable branch of the flow curve, so the $\mathrm{Re}$ trends in Figs.~6 and 7 of Ref.~\onlinecite{balan2023} are trends with gap width (in elastic lengths) at one, rather special, apparent shear rate. 

(ii)~For $a = \pm 1$, Eqs.~\eqref{eq:1} are linear in $v,\sigma$, and the start-up of plane Couette flow at arbitrary $\mathrm{Re}$ has the exact solution of Ref.~\onlinecite{christov2013}, Eqs.~(42)--(43), with $\tau = 1/\mathrm{Re}$, $\alpha = \kappa/\mathrm{Re}$, and time rescaled by $1/\mathrm{Re}$. Tanner's half-space solution, like other classical solutions~\cite{waters_king1970,denn_porteous1971}, likewise includes unsteady inertia. The statement in Ref.~\onlinecite{balan2023} that the earlier analytic solutions were obtained ``without investigating explicitly the influence of the Reynolds number (most of the solutions being obtained in the limit $\mathrm{Re} \to 0$)'' is therefore incorrect. Nor is the description of Tanner's exact (Laplace-inversion) solution as ``the first numerical solution.'' In fact, Morrison~\cite{morrison1956} had solved the same equation [i.e., Eq.~(8) of Ref.~\onlinecite{balan2023}] exactly for the impulsive start-up condition six years before Tanner~\cite{tanner1962}.

(iii)~Four of the seven works cited in Ref.~\onlinecite{balan2023} as sources of ``analytic viscoelastic solutions for the Oldroyd-B model'' are erroneous, as is a fifth that it cites for the second-order fluid. Ref.~21 of Ref.~\onlinecite{balan2023} was already corrected in \emph{this} journal;\cite{christov_jordan2009} Refs.~18 and 19 of Ref.~\onlinecite{balan2023} have been shown to be erroneous,\cite{christov_jordan2012,christov2015} as have Ref.~20 of Ref.~\onlinecite{balan2023} and its erratum;\cite{christov2010,christov2011} and Ref.~24 of Ref.~\onlinecite{balan2023} introduced the eigenfunction-expansion treatment of start-up flows that Ref.~\onlinecite{christov2013} corrects. The error stems from mishandling the impulsive start-up~\cite{christov2010}. Tanner's~\cite{tanner1962} use of the Laplace transform in $t$ on the consistently posed IBVP avoids it.

(iv)~The second panel of Fig.~4(a) of Ref.~\onlinecite{balan2023}, labeled $\kappa = 0.2$, repeats the $\kappa = 0.4$ curve of the first, as digitizing both panels shows~\cite{christov_code2026}.

\appendix*
\section{What problem does the IBVP of Ref.~1 solve?}

With $S = \sigma - \kappa\partial_x v$ and $a = \pm1$, $N$ decouples. Eqs.~\eqref{eq:1} read $\mathrm{Re}\partial_t v = \partial_x S + \kappa\partial_x^2 v$ and $\partial_t S + S = (1-\kappa)\partial_x v$ on $0 < x < x_0$. The \emph{consistently posed start-up} has $v = S = 0$ for $t < 0$, $v(x_0, t) = v_0 H(t)$, and $v(0,t) = 0$. Taking the Laplace transform in $t$ (parameter $p$, overbars) with the data of IBVP (9)--(10), for which $S(x,0) = -\kappa v_0\delta(x-x_0)$, and noting that $v(x,0) = 0$ for $0 < x < x_0$,
\begin{subequations}\begin{align}
    (1+p)\bar S &= (1-\kappa)\partial_x \bar v - \kappa v_0 \delta(x-x_0), \label{eq:A1a}\\ 
    \mathrm{Re}\,p\,\bar v &= \partial_x \bar S + \kappa\partial_x^2 \bar v . \label{eq:A1b}
\end{align}\end{subequations}
Eliminating $\bar S$ between Eqs.~\eqref{eq:A1a} and \eqref{eq:A1b},
\begin{equation}
    \mathrm{Re}\,p\,\bar v = \left(\frac{1+\kappa p}{1+p}\right) \partial_x^2 \bar v - \frac{\kappa v_0}{1+p}\delta'(x-x_0). \label{eq:A2}
\end{equation}
Let $\bar w = \bar v - [\kappa v_0/(1+\kappa p)]H(x-x_0)$, which coincides with $\bar v$ on
$0<x<x_0$ and is continuous at $x_0$, since the $\delta'$ jumps $\bar v$ by that amount there. It satisfies Eq.~\eqref{eq:A2} \emph{without} its source there, which is the Laplace-domain equation of the consistently posed problem (Ref.~\onlinecite{tanner1962}; Ref.~\onlinecite{christov_jordan2009}, Eq.~(9)). Then $\bar w(0,p)=\bar v(0,p)=0$ and, from the plate IC $\bar v(x_0,p)=v_0/p$ with the jump subtracted,
\begin{equation}
    \bar w(x_0, p) = v_0\left(\frac{1}{p} - \frac{\kappa}{1+\kappa p}\right)
    = \mathcal{L}\left\{v_0\left(1 - \mathrm{e}^{-t/\kappa}\right)\right\}, \label{eq:A3}
\end{equation}
and it follows that the velocity field of IBVP (9)--(10) of Ref.~\onlinecite{balan2023} is that of the start-up problem with plate velocity $v_0(1 - \mathrm{e}^{-t/\kappa})$, i.e., $v_0(1 - \mathrm{e}^{-t/\lambda_2})$ in dimensional form. In closed form, it is Eq.~(7) of Ref.~\onlinecite{christov_jordan2009} with the term in $\alpha_t$, the one driven by the $\delta(t)$ of the start-up jump, deleted---the same term that the corrected papers drop. 

One limit checks this: as $\kappa \to 0$, both the fictitious stress and the ramp vanish, as they must, since the error requires a retardation time. At the other extreme, $\kappa = 1$, the ramp time is the viscous time and the discrepancy is largest. Numerically, finite differences reproduce the ramped-plate solution to better than $10^{-4}$ across a range of $\kappa$. For $a = 0$, the same fictitious stress is present, but it enters Eq.~\eqref{eq:1b} through $S\partial_x v$, so no closed-form reinterpretation exists, and it leads to the grid-dependent transient shown in Fig.~\ref{fig:1}(c).

\begin{acknowledgments}
I.C.C.\ identified the error in the initial conditions of Ref.~\onlinecite{balan2023}, connected it to the prior corrections~\cite{christov_christov2010,christov_jordan2009,christov2013,christov2015,christov_jordan2012,christov2010,christov2011}, conducted the literature survey, and interpreted the results. I.C.C.\ employed Claude (Anthropic; models Fable 5.1 via claude.ai, and Opus 5 via Claude Code, September 2026) alternating as a coding and calculation assistant or adversary (asked to find errors or inconsistencies in I.C.C.'s logic). Under the author's direction, Claude Code (i)~adapted the author's existing \textsc{Matlab} and \textsc{Mathematica} codes into Python and validated them; (ii)~wrote scripts to digitize curves from Figs.~2, 3(a), and 4(a) of Ref.~\onlinecite{balan2023} for comparison with the exact solution; (iii)~set up the grid-refinement computations; and (iv)~helped set up the open-source code repository. All computational results were checked by the author, who wrote the Comment and is responsible for its content.
\end{acknowledgments}

\section*{Author declarations}
\subsection*{Conflict of Interest}
The author has no conflicts to disclose.

\section*{Data availability}
The data that support the findings of this study are openly available in the archived repository~\cite{christov_code2026}, notebook \texttt{balan\_2023\_startup\_couette.ipynb} of \url{https://github.com/ichristov/viscoelastic-startup}, which reproduces every figure and number in this Comment.

\newpage

\bibliographystyle{aipnum4-2}
\bibliography{balan_comment}

%aipnum4-2.bst 2019-01-14 (MD) hand-edited version of apsrev4-1.bst
\begin{thebibliography}{21}%
\makeatletter
\providecommand \@ifxundefined [1]{%
 \@ifx{#1\undefined}
}%
\providecommand \@ifnum [1]{%
 \ifnum #1\expandafter \@firstoftwo
 \else \expandafter \@secondoftwo
 \fi
}%
\providecommand \@ifx [1]{%
 \ifx #1\expandafter \@firstoftwo
 \else \expandafter \@secondoftwo
 \fi
}%
\providecommand \natexlab [1]{#1}%
\providecommand \enquote  [1]{``#1''}%
\providecommand \bibnamefont  [1]{#1}%
\providecommand \bibfnamefont [1]{#1}%
\providecommand \citenamefont [1]{#1}%
\providecommand \href@noop [0]{\@secondoftwo}%
\providecommand \href [0]{\begingroup \@sanitize@url \@href}%
\providecommand \@href[1]{\@@startlink{#1}\@@href}%
\providecommand \@@href[1]{\endgroup#1\@@endlink}%
\providecommand \@sanitize@url [0]{\catcode `\\12\catcode `\$12\catcode
  `\&12\catcode `\#12\catcode `\^12\catcode `\_12\catcode `\%12\relax}%
\providecommand \@@startlink[1]{}%
\providecommand \@@endlink[0]{}%
\providecommand \url  [0]{\begingroup\@sanitize@url \@url }%
\providecommand \@url [1]{\endgroup\@href {#1}{\urlprefix }}%
\providecommand \urlprefix  [0]{URL }%
\providecommand \Eprint [0]{\href }%
\providecommand \doibase [0]{https://doi.org/}%
\providecommand \selectlanguage [0]{\@gobble}%
\providecommand \bibinfo  [0]{\@secondoftwo}%
\providecommand \bibfield  [0]{\@secondoftwo}%
\providecommand \translation [1]{[#1]}%
\providecommand \BibitemOpen [0]{}%
\providecommand \bibitemStop [0]{}%
\providecommand \bibitemNoStop [0]{.\EOS\space}%
\providecommand \EOS [0]{\spacefactor3000\relax}%
\providecommand \BibitemShut  [1]{\csname bibitem#1\endcsname}%
\let\auto@bib@innerbib\@empty
%</preamble>
\bibitem [{\citenamefont {Balan}(2023)}]{balan2023}%
  \BibitemOpen
  \bibfield  {author} {\bibinfo {author} {\bibfnamefont {C.}~\bibnamefont
  {Balan}},\ }\href {https://doi.org/10.1063/5.0173510} {\bibfield  {journal}
  {\bibinfo  {journal} {Phys. Fluids}\ }\textbf {\bibinfo {volume} {35}},\
  \bibinfo {pages} {113108} (\bibinfo {year} {2023})}\BibitemShut {NoStop}%
\bibitem [{\citenamefont {Gordon}\ and\ \citenamefont
  {Schowalter}(1972)}]{gordon_schowalter1972}%
  \BibitemOpen
  \bibfield  {author} {\bibinfo {author} {\bibfnamefont {R.~J.}\ \bibnamefont
  {Gordon}}\ and\ \bibinfo {author} {\bibfnamefont {W.~R.}\ \bibnamefont
  {Schowalter}},\ }\href {https://doi.org/10.1122/1.549256} {\bibfield
  {journal} {\bibinfo  {journal} {Trans. Soc. Rheol.}\ }\textbf {\bibinfo
  {volume} {16}},\ \bibinfo {pages} {79} (\bibinfo {year} {1972})}\BibitemShut
  {NoStop}%
\bibitem [{\citenamefont {Johnson}\ and\ \citenamefont
  {Segalman}(1977)}]{johnson_segalman1977}%
  \BibitemOpen
  \bibfield  {author} {\bibinfo {author} {\bibfnamefont {M.~W.}\ \bibnamefont
  {Johnson}}\ and\ \bibinfo {author} {\bibfnamefont {D.}~\bibnamefont
  {Segalman}},\ }\href {https://doi.org/10.1016/0377-0257(77)80003-7}
  {\bibfield  {journal} {\bibinfo  {journal} {J. Non-Newtonian Fluid Mech.}\
  }\textbf {\bibinfo {volume} {2}},\ \bibinfo {pages} {255} (\bibinfo {year}
  {1977})}\BibitemShut {NoStop}%
\bibitem [{\citenamefont {Tanner}(1962)}]{tanner1962}%
  \BibitemOpen
  \bibfield  {author} {\bibinfo {author} {\bibfnamefont {R.~I.}\ \bibnamefont
  {Tanner}},\ }\href {https://doi.org/10.1007/BF01595580} {\bibfield  {journal}
  {\bibinfo  {journal} {Z. Angew. Math. Phys.}\ }\textbf {\bibinfo {volume}
  {13}},\ \bibinfo {pages} {573} (\bibinfo {year} {1962})}\BibitemShut
  {NoStop}%
\bibitem [{\citenamefont {Christov}\ and\ \citenamefont
  {Christov}(2010)}]{christov_christov2010}%
  \BibitemOpen
  \bibfield  {author} {\bibinfo {author} {\bibfnamefont {I.~C.}\ \bibnamefont
  {Christov}}\ and\ \bibinfo {author} {\bibfnamefont {C.~I.}\ \bibnamefont
  {Christov}},\ }\href {https://doi.org/10.1007/s00707-010-0300-2} {\bibfield
  {journal} {\bibinfo  {journal} {Acta Mech.}\ }\textbf {\bibinfo {volume}
  {215}},\ \bibinfo {pages} {25} (\bibinfo {year} {2010})}\BibitemShut
  {NoStop}%
\bibitem [{\citenamefont {Barenblatt}, \citenamefont {Zheltov},\ and\
  \citenamefont {Kochina}(1960)}]{barenblatt1960}%
  \BibitemOpen
  \bibfield  {author} {\bibinfo {author} {\bibfnamefont {G.~I.}\ \bibnamefont
  {Barenblatt}}, \bibinfo {author} {\bibfnamefont {I.~P.}\ \bibnamefont
  {Zheltov}},\ and\ \bibinfo {author} {\bibfnamefont {I.~N.}\ \bibnamefont
  {Kochina}},\ }\href {https://doi.org/10.1016/0021-8928(60)90107-6} {\bibfield
   {journal} {\bibinfo  {journal} {J. Appl. Math. Mech.}\ }\textbf {\bibinfo
  {volume} {24}},\ \bibinfo {pages} {1286} (\bibinfo {year}
  {1960})}\BibitemShut {NoStop}%
\bibitem [{\citenamefont {Oldroyd}(1950)}]{oldroyd1950}%
  \BibitemOpen
  \bibfield  {author} {\bibinfo {author} {\bibfnamefont {J.~G.}\ \bibnamefont
  {Oldroyd}},\ }\href {https://doi.org/10.1098/rspa.1950.0035} {\bibfield
  {journal} {\bibinfo  {journal} {Proc. R. Soc. Lond. A}\ }\textbf {\bibinfo
  {volume} {200}},\ \bibinfo {pages} {523} (\bibinfo {year}
  {1950})}\BibitemShut {NoStop}%
\bibitem [{\citenamefont {Christov}\ and\ \citenamefont
  {Jordan}(2009)}]{christov_jordan2009}%
  \BibitemOpen
  \bibfield  {author} {\bibinfo {author} {\bibfnamefont {C.~I.}\ \bibnamefont
  {Christov}}\ and\ \bibinfo {author} {\bibfnamefont {P.~M.}\ \bibnamefont
  {Jordan}},\ }\href {https://doi.org/10.1063/1.3126503} {\bibfield  {journal}
  {\bibinfo  {journal} {Phys. Fluids}\ }\textbf {\bibinfo {volume} {21}},\
  \bibinfo {pages} {069101} (\bibinfo {year} {2009})}\BibitemShut {NoStop}%
\bibitem [{\citenamefont {Christov}(2013)}]{christov2013}%
  \BibitemOpen
  \bibfield  {author} {\bibinfo {author} {\bibfnamefont {I.~C.}\ \bibnamefont
  {Christov}},\ }\href {https://doi.org/10.1016/j.mechrescom.2013.05.005}
  {\bibfield  {journal} {\bibinfo  {journal} {Mech. Res. Commun.}\ }\textbf
  {\bibinfo {volume} {51}},\ \bibinfo {pages} {86} (\bibinfo {year}
  {2013})}\BibitemShut {NoStop}%
\bibitem [{\citenamefont {Christov}(2026)}]{christov_code2026}%
  \BibitemOpen
  \bibfield  {author} {\bibinfo {author} {\bibfnamefont {I.~C.}\ \bibnamefont
  {Christov}},\ }\href {https://doi.org/10.5281/zenodo.23005137} {\enquote
  {\bibinfo {title} {{\texttt{viscoelastic-startup}: Verification and
  reproducibility of exact unidirectional solutions for start-up flows of
  viscoelastic fluids}},}\ }\bibinfo {howpublished} {Zenodo} (\bibinfo {year}
  {2026}),\ \bibinfo {note} {version v1.1.0,
  \href{https://doi.org/10.5281/zenodo.23005137}{doi:10.5281/zenodo.23005137}}\BibitemShut
  {NoStop}%
\bibitem [{\citenamefont {Divoux}\ \emph {et~al.}(2016)\citenamefont {Divoux},
  \citenamefont {Fardin}, \citenamefont {Manneville},\ and\ \citenamefont
  {Lerouge}}]{divoux2016}%
  \BibitemOpen
  \bibfield  {author} {\bibinfo {author} {\bibfnamefont {T.}~\bibnamefont
  {Divoux}}, \bibinfo {author} {\bibfnamefont {M.~A.}\ \bibnamefont {Fardin}},
  \bibinfo {author} {\bibfnamefont {S.}~\bibnamefont {Manneville}},\ and\
  \bibinfo {author} {\bibfnamefont {S.}~\bibnamefont {Lerouge}},\ }\href
  {https://doi.org/10.1146/annurev-fluid-122414-034416} {\bibfield  {journal}
  {\bibinfo  {journal} {Annu. Rev. Fluid Mech.}\ }\textbf {\bibinfo {volume}
  {48}},\ \bibinfo {pages} {81} (\bibinfo {year} {2016})}\BibitemShut {NoStop}%
\bibitem [{\citenamefont {Stephanou}(2023)}]{stephanou2023}%
  \BibitemOpen
  \bibfield  {author} {\bibinfo {author} {\bibfnamefont {P.~S.}\ \bibnamefont
  {Stephanou}},\ }\href {https://doi.org/10.1016/j.jnnfm.2022.104966}
  {\bibfield  {journal} {\bibinfo  {journal} {J. Non-Newtonian Fluid Mech.}\
  }\textbf {\bibinfo {volume} {312}},\ \bibinfo {pages} {104966} (\bibinfo
  {year} {2023})}\BibitemShut {NoStop}%
\bibitem [{\citenamefont {Georgiou}\ and\ \citenamefont
  {Vlassopoulos}(1998)}]{georgiou1998}%
  \BibitemOpen
  \bibfield  {author} {\bibinfo {author} {\bibfnamefont {G.~C.}\ \bibnamefont
  {Georgiou}}\ and\ \bibinfo {author} {\bibfnamefont {D.}~\bibnamefont
  {Vlassopoulos}},\ }\href {https://doi.org/10.1016/S0377-0257(97)00078-5}
  {\bibfield  {journal} {\bibinfo  {journal} {J. Non-Newtonian Fluid Mech.}\
  }\textbf {\bibinfo {volume} {75}},\ \bibinfo {pages} {77} (\bibinfo {year}
  {1998})}\BibitemShut {NoStop}%
\bibitem [{\citenamefont {Joseph}(1986)}]{joseph1986}%
  \BibitemOpen
  \bibfield  {author} {\bibinfo {author} {\bibfnamefont {D.~D.}\ \bibnamefont
  {Joseph}},\ }\href {https://doi.org/10.1016/0377-0257(86)80051-9} {\bibfield
  {journal} {\bibinfo  {journal} {J. Non-Newtonian Fluid Mech.}\ }\textbf
  {\bibinfo {volume} {19}},\ \bibinfo {pages} {237} (\bibinfo {year}
  {1986})}\BibitemShut {NoStop}%
\bibitem [{\citenamefont {Waters}\ and\ \citenamefont
  {King}(1970)}]{waters_king1970}%
  \BibitemOpen
  \bibfield  {author} {\bibinfo {author} {\bibfnamefont {N.~D.}\ \bibnamefont
  {Waters}}\ and\ \bibinfo {author} {\bibfnamefont {M.~J.}\ \bibnamefont
  {King}},\ }\href {https://doi.org/10.1007/BF01975401} {\bibfield  {journal}
  {\bibinfo  {journal} {Rheol. Acta}\ }\textbf {\bibinfo {volume} {9}},\
  \bibinfo {pages} {345} (\bibinfo {year} {1970})}\BibitemShut {NoStop}%
\bibitem [{\citenamefont {Denn}\ and\ \citenamefont
  {Porteous}(1971)}]{denn_porteous1971}%
  \BibitemOpen
  \bibfield  {author} {\bibinfo {author} {\bibfnamefont {M.~M.}\ \bibnamefont
  {Denn}}\ and\ \bibinfo {author} {\bibfnamefont {K.~C.}\ \bibnamefont
  {Porteous}},\ }\href {https://doi.org/10.1016/0300-9467(71)85007-4}
  {\bibfield  {journal} {\bibinfo  {journal} {Chem. Eng. J.}\ }\textbf
  {\bibinfo {volume} {2}},\ \bibinfo {pages} {280} (\bibinfo {year}
  {1971})}\BibitemShut {NoStop}%
\bibitem [{\citenamefont {Morrison}(1956)}]{morrison1956}%
  \BibitemOpen
  \bibfield  {author} {\bibinfo {author} {\bibfnamefont {J.~A.}\ \bibnamefont
  {Morrison}},\ }\href {https://doi.org/10.1090/qam/78848} {\bibfield
  {journal} {\bibinfo  {journal} {Quart. Appl. Math.}\ }\textbf {\bibinfo
  {volume} {14}},\ \bibinfo {pages} {153} (\bibinfo {year} {1956})}\BibitemShut
  {NoStop}%
\bibitem [{\citenamefont {Christov}\ and\ \citenamefont
  {Jordan}(2012)}]{christov_jordan2012}%
  \BibitemOpen
  \bibfield  {author} {\bibinfo {author} {\bibfnamefont {I.~C.}\ \bibnamefont
  {Christov}}\ and\ \bibinfo {author} {\bibfnamefont {P.~M.}\ \bibnamefont
  {Jordan}},\ }\href {https://doi.org/10.1016/j.ijengsci.2011.10.012}
  {\bibfield  {journal} {\bibinfo  {journal} {Int. J. Eng. Sci.}\ }\textbf
  {\bibinfo {volume} {51}},\ \bibinfo {pages} {326} (\bibinfo {year}
  {2012})}\BibitemShut {NoStop}%
\bibitem [{\citenamefont {Christov}(2015)}]{christov2015}%
  \BibitemOpen
  \bibfield  {author} {\bibinfo {author} {\bibfnamefont {I.~C.}\ \bibnamefont
  {Christov}},\ }\href {https://doi.org/10.1139/cjp-2015-0374} {\bibfield
  {journal} {\bibinfo  {journal} {Can. J. Phys.}\ }\textbf {\bibinfo {volume}
  {93}},\ \bibinfo {pages} {1651} (\bibinfo {year} {2015})}\BibitemShut
  {NoStop}%
\bibitem [{\citenamefont {Christov}(2010)}]{christov2010}%
  \BibitemOpen
  \bibfield  {author} {\bibinfo {author} {\bibfnamefont {I.~C.}\ \bibnamefont
  {Christov}},\ }\href {https://doi.org/10.1016/j.mechrescom.2010.09.006}
  {\bibfield  {journal} {\bibinfo  {journal} {Mech. Res. Commun.}\ }\textbf
  {\bibinfo {volume} {37}},\ \bibinfo {pages} {717} (\bibinfo {year}
  {2010})}\BibitemShut {NoStop}%
\bibitem [{\citenamefont {Christov}(2011)}]{christov2011}%
  \BibitemOpen
  \bibfield  {author} {\bibinfo {author} {\bibfnamefont {I.~C.}\ \bibnamefont
  {Christov}},\ }\href {https://doi.org/10.1016/j.nonrwa.2011.06.025}
  {\bibfield  {journal} {\bibinfo  {journal} {Nonlinear Anal. Real World
  Appl.}\ }\textbf {\bibinfo {volume} {12}},\ \bibinfo {pages} {3687} (\bibinfo
  {year} {2011})}\BibitemShut {NoStop}%
\end{thebibliography}%

\end{document}